\documentclass[iop,revtex4]{emulateapj}   
\usepackage[titletoc,title]{appendix}
\usepackage{lscape}
\usepackage{graphicx,natbib,url,twoopt}

\usepackage{hyperref}               
\usepackage{blkarray}
\usepackage{url}
\usepackage{multirow}
\usepackage{bm}

\usepackage{amsmath} 
\usepackage[varg]{txfonts}

\bibpunct{(}{)}{;}{a}{}{,}    

\received{receipt date}
\revised{revision date}
\accepted{acceptance date}
\begin{document}  

\title{
Reconciling the predictions of microlensing analysis with radial velocity measurements for OGLE-2011-BLG-0417.
}

\author{E. Bachelet$^{1}$, J.-P. Beaulieu$^{2,3}$, I. Boisse$^{4}$, A.Santerne$^{4}$ and R.A. Street$^{1}$.}

\affil{$^1$Las Cumbres Observatory, 6740 Cortona Drive, Suite 102, Goleta, CA 93117 USA}
\affil{$^2$ \quad School of Physical Sciences, University of Tasmania, Private Bag 37 Hobart, Tasmania 7001 Australia; jeanphilippe.beaulieu@utas.edu.au}
\affil{$^3$ \quad Sorbonne Universit\' es, UPMC Universit\' e Paris 6 et CNRS, UMR 7095, Institut d'Astrophysique de Paris, 98http://www.plt.axvline/ bis bd Arago, 75014 Paris, France; beaulieu@iap.fr}
\affil{$^4$ \quad Aix Marseille Univ, CNRS, CNES, LAM, Marseille, France}

\begin{abstract}
Microlensing is able to reveal multiple body systems located several kilo-parsec away from the Earth. Since it does not require the measurement of light from the lens, microlensing is sensitive to a range of objects from free-floating planets to stellar black holes. But if the lens emits enough light, the microlensing model predictions can be tested with high-resolution imaging and/or radial velocity methods. Such follow-up was done for the microlensing event OGLE-2011-BLG-0417, which was expected to be a close by ($\le$ 1 kpc), low-mass ($\sim 0.8 M_\odot$) binary star with a period $P\sim 1.4$ yr. The spectroscopic follow-up observations conducted with the VLT did not measure any variation in the radial velocity, in strong contradiction with the published microlensing model. In the present work, we remodel this event and find a simpler model in agreement with all the available measurements, including the recent GAIA DR2 parallax constraints. We also present a new way to distinguish degenerate models by using the GAIA DR2 proper motions. This work stresses the importance of thorough microlensing modeling, especially with the horizon of the {\it WFIRST} and the {\it Euclid} microlensing space missions.
\end{abstract}
\section{Introduction}     \label{sec:introduction}
The gravitational microlensing technique \citep{Paczynski1986} is a unique tool to explore stellar and planetary companions over a very wide range of masses and orbits along the line of sight towards the Galactic Center \citep{Gould1992, Beaulieu2006, Cassan2012,Clanton2016,Suzuki2016,Penny2016}.
Usually, accurate mass ratios and projected separations in units of the Einstein ring radius are extracted by modeling the microlensing light curve. One mass-distance relation for the lens is generally obtained in the case of binary lens events, from the measurements of finite-source effects. Another lens mass-distance relation can be derived from the parallax effects along with the lens light detection and/or a Bayesian analysis with a galactic model (see \citet{Beaulieu2018} and references therein). This is then sufficient to estimate the physical parameters (mass, distance, projected orbital separation) of the lens system.
In rare cases where the lens is bright enough to be detected with good contrast with the source, it is possible to perform radial velocity follow-up observations to test and refine the microlensing predictions \citep{Yee2016}.

OGLE-2011-BLG-0417, originally presented by \citet{Shin2012} (S12 hereafter), is a 
microlensing event with strong features of a binary lens. The authors conducted a deep analysis and found a full Keplerian solution for the lens system. The lens was expected to be close by (i.e $D_l\sim0.9$ kpc) and composed of low-mass binary stars (i.e $M_{tot}\sim 0.7 ~\rm{M_\odot}$). Moreover, assuming partial extinction, S12 suggested that "the blended light...comes very likely from the lens itself, implying that the lens system can be directly observed". 

Based on the S12 solution, \citet{Gould2013} (G13 afterwards) did not repeat the modeling but derived radial velocity properties of the lens system with $K=6.31 \pm0.34 ~\rm{km}~ \rm{ s^{-1}}$ and $P=1.423 \pm 0.113 ~\rm{yr}$. Therefore, G13 concluded that the radial velocity signal "should be precisely measurable despite the fact that the lens primary is relatively faint ($I=16.3$ mag, $V=18.2$ mag)."

Following the prediction of G13, \citet{Boisse2015} (B15 hereafter) observed the target with the UVES spectrograph \citep{Dekker2000} mounted on the VLT.
From the high-precision radial velocity measurements (i.e RMS=94 $\rm{m/s}$) based on the 10 spectra taken over a year, they found that there was no measurable modulation in the radial velocity. This result contradicted the predictions of G13 and B15 concluded that the microlensing scenario of G13 was highly unlikely, given their new data (with a probability $P\le10^{-7}$). This raised doubts about the microlensing predictions.

In addition to the B15 investigations, \citet{Santerne2016} (S16 in the following) obtained supplementary data on this target. The authors secured 
a near-infrared spectrum using the ARCoIRIS spectrograph at CTIO \citep{Schlawin2014}. They also obtained near-infrared
high-resolution images using the NIRC2 adaptive optics instrument at the Keck II telescope. Their results indicate that the source is a giant star in the Galactic Bulge and the lens is located at a distance of $D_l\sim1$ kpc. They confirm previous results from B15 and 
conclude that the microlensing model is probably wrong.

In this work, we reanalyze the microlensing event OGLE-2011-BLG-0417. We present the observations in the Section~\ref{sec:obs}. In the Section~\ref{sec:modeling}, we first 
study in details the results of S12 and find clues that corroborate the conclusions of B15 and S16. We then present our modeling process and the new results. We finally conclude in the Section~\ref{sec:conclusion}.
\section{Microlensing Observations} \label{sec:obs}
The microlensing event OGLE-2011-BLG-0417 ($\alpha={\rm 17^h34^m33.12^s},\delta={\rm -27^\circ 06^m 39.3^s}$(J2000)) was 
alerted by the Optical Gravitational Lens Experiment (OGLE) \citep{Udalski2015a}. Because it is a bright event (i.e $I\le15.7$), several teams conducted follow-up observations based on the OGLE alert and obtained data during the magnification peak. In this work, 
we use the same datasets described in S12 that we summarize in the Table~\ref{tab:sumobservations}. OGLE, PLANET, MiNDSTEp and RoboNet data were obtained by the 
difference image analysis (DIA) technique. RoboNet used {\tt DanDIA} \citep{Bramich2008,Bramich2013}, 
OGLE used their own implementation of DIA \citep{Udalski2015a} while PLANET and MiNDSTEp used pySIS \citep{Albrow2009}.
A special pipeline based on DoPHOT \citep{Schechter1993} was used to process $\mu$FUN data.

It is common practice in microlensing to rescale the photometric uncertainties using (in mag unit):
\begin{equation}
\sigma' = k\sqrt{\sigma^2+e_{min}^2}~,
\end{equation}
where $\sigma'$ are the rescaled uncertainties, $\sigma$ are the original uncertainties, $k$ a parameter to adjust low-magnification 
uncertainties and $e_{min}$ is used to adjust uncertainties at high-magnification. To compare this work with S12, 
we use the same rescaling parameters as S12 that we report in Table~\ref{tab:sumobservations}.

This event was also in the footprint of the VISTA Variables in the Via Lactea (VVV) survey \citep{Minniti2010}. This dataset was not used by S12.
The $K$-band data obtained by the VVV survey were re-reduced using pySIS. We then align the pySIS reduction to an independent 
VVV catalog \citep{Beaulieu2016} by adding a 0.7 mag offset to the pySIS lightcurve. However, 
the target is so bright in the near-infrared that it is close to saturation during the highest amplification. The photometry quality 
during the central caustic approach is therefore low. Moreover, the number of data points near the peak of the lensing event is small ($\le10$) and the data do not cover the caustic crossing.
For these reasons and in order to make a like-for-like comparison with S12, we opt not to include the VVV data in our modeling process, and instead used it afterwards to obtain the source brightness in $K$ band.
\begin{table*}
  \footnotesize
  
  \centering
  \begin{tabular}{lccccccc}
    & \\
     \hline\hline
Name&Collaboration&Aperture(m)&Filter&Code&$N_{data}$&k&$e_{min}$ \\
    \hline
      & \\
 OGLE\_I&OGLE&1.3&I&Wo\'{z}niak&1481&2.740&0.0005 \\
 OGLE\_V&OGLE&1.3&V&Wo\'{z}niak&32&1.300&0.0005 \\
 CTIO\_I&$\rm{\mu FUN}$&1.3&I&DoPHOT&18&1.440&0.0005 \\
 CTIO\_V&$\rm{\mu FUN}$&1.3&V&DoPHOT&5&0.597&0.0005 \\
 Auckland\_R&$\rm{\mu FUN}$&0.4&Red&DoPHOT&9&0.635&0.0005 \\
 FCO\_U&$\rm{\mu FUN}$&0.4&U&DoPHOT&72&2.470&0.0005 \\
 Kumeu\_R&$\rm{\mu FUN}$&0.4&Red&DoPHOT&15&0.773&0.0005 \\
 OPD\_I&$\rm{\mu FUN}$&0.6&I&DoPHOT&520&2.325&0.0005 \\
 Canopus\_I&PLANET&1.0&I&pySIS&122&1.320&0.0005 \\
 SAAO\_I&PLANET&1.0&I&pySIS&29&7.035&0.0005 \\
 Danish\_I&MiNDSTEp&1.5&I&pySIS&118&3.350&0.0005 \\
 FTN\_i&RoboNet&2.0&SDSS-i'&DANDIA&140&3.860&0.0005 \\
 LT\_i&RoboNet&2.0&SDSS-i'&DANDIA&73&2.250&0.0005 \\
 \hline
 VVV\_K&VVV&4.0&K&pySIS&148&1.0&0.05 \\
   
    \hline
  \end{tabular}

  \caption{Summary of observations. Instrumental magnitudes of OGLE are for the Cousin I filter and close to the Johnson V filter \citep{Udalski2015a}. The $\mu FUN$ filters are : Bessell V, Cousins I, a large band R and U is unfiltered (modulo the CCD quantum efficiency curve). The I-band of PLANET is Cousin I \citep{Albrow1998}, while RoboNet filters are SDSS-i'. MiNDSTEp I filter is Gunn i, close to the standard system according to \citet{Skowron2016}.}
  \label{tab:sumobservations}
\end{table*}

\section{Revisiting the Microlensing Modeling} \label{sec:modeling}
\subsection{Review of previous results} \label{sec:history}
As described in S12, the static binary microlensing magnification is computed with seven parameters. $t_0$ is the time of the minimum impact parameters $u_0$ (relative to origin of the system, the center of mass in this work), $t_E$ is the angular Einstein radius crossing time, 
$\rho$ is the normalized angular source radius, $s$ is the normalized 
projected separation, $q$ is the mass ratio between the two bodies lens component and $\alpha$ the lens/source trajectory angle relative to the binary axes. Finally, the flux measurements from each telescope are used to derive two extra linear parameters : the source flux $f_s$ and the blend flux $f_b$. For a more complete description of the problem, readers can refer to \citet{Mao2012} and \citet{Gaudi2012}. 

S12 modeled this event and found that the anomalies were due to source star's crossings over a caustic close to the primary. This geometry might be sensitive to the close/wide degeneracy \citep{Dominik1999,Bozza2000}, where the central caustic shape is similar if the companion is inside (close model) or beyond (wide model) the Einstein ring. However, S12 only reported a close model ($s\le 1$) in their analysis. S12 used the grid search method to find this solution, whereby they explored the $(\log(s),\log(q))$ parameter space on a grid, with all other parameters optimized with a downhill approach, generally using a Monte Carlo Markov Chain algorithm \citep{Dong2006}. Unfortunately, there is so far no method available to determine whether all possible minima have been discovered and hence no way to ensure that the optimum solutions has been identified (in a finite time). The grid search generally performs well but is sensitive to the grid sampling and we are aware that it failed at least in two occasions. For OGLE-2008-BLG-513, \citet{Yee2011} found a model where the lens is composed of a super-Jupiter orbiting an M-dwarf, but further modeling work by \citet{Jeong2015} found that a binary lens composed of similar mass components better described the observed data. More recently, \citet{Han2016} revisited the event OGLE-2013-BLG-0723 and concluded that it resulted from a lens composed of two low-mass stars, rather than, as originally published by \citet{Udalski2015b}, a triple-body lens composed of a Venus-mass planet orbiting a brown-dwarf in a binary system.

By examining in more detail the microlensing models of S12 (their Figure 2), it is interesting to note a bump in the residuals of their static model around ${\rm HJD=2455730}$. This is due to the passage of the source trajectory close to a peripheral caustic (see their Figure 3). Including microlensing parallax \citep{Hardy1995,Smith2003,Gould2004} and the orbital motion of the lens \citep{Dominik1998,Ioka1999,Albrow2000} in their static model allowed the necessary flexibility for the source trajectory to avoid the passage close to this peripheral caustic, dramatically improving the $\chi^2$ since this deviation is not supported by the data. Moreover, the orbital motion effect modify the projected separation of the lens components, inducing a modification of the caustics pattern. In the analysis of S12, the rate of binary separation change was considerable (${{ds_\perp}\over{dt}} = 1.314 \pm 0.023 ~{\rm yr^{-1}}$). The triangular caustic is then very distant from the trajectory at the date ${\rm HJD=2455730}$, improving again the $\chi^2$ in their best model (as can be seen in their Table 3). We note that the close/wide degeneracy implies that a competitive wide model may exist, for which the source trajectory would not intersect this peripheral caustic, potentially leading to a better model.  

The presence of the "bump" at $\rm{HJD}=2455730$ and the absence of a static wide-binary solution in S12 analysis might indicate that their model corresponds to a local minima, raising the need for further investigation. However, we note that G13 favored an alternative conclusion: "Indeed, the fact that this blended-light point sits right on the "disk main sequence" or "reddening track" at roughly the position expected for the primary component of the lens is already an indication that the microlensing solution is basically correct."   

\subsection{Description} \label{sec:modeldescription}
We reanalyzed this microlensing event using the pyLIMA\footnote{https://github.com/ebachelet/pyLIMA} software \citep{Bachelet2017}. A complete description of the binary fitting 
with pyLIMA will be given in Bachelet (2018, in prep).
Whereas S12 considered limb-darkening effects in their models, in this work, we consider a simpler model,  the 
Uniform Source Binary Lens (USBL) \citep{Bozza2012}. The motivation for this is that caustic crossings are covered by a small number of data points (i.e $\le10$ data points) and they were taken in similar passbands (I and SDSS-i'). In this case, the limb-darkening is weakly constrained. The results of our analysis show that this assumption 
is verified.

Instead of using the same grid approach as S12, we run a global search on all parameters using the differential evolution method. See \citet{Storn1997} and \citet{Bachelet2017} for details. However, due to the dramatic difference in the caustic topology between the wide and close binary regimes \citep{Erdl1993,Dominik1999,Bozza2000,Cassan2008}, we split the parameter space into the close ($s<1$) and the wide ($s>1$) regimes. This helps to avoid the potential pitfall, described in the previous section, of missing degenerate solutions.

To obtain a more comprehensive picture, each minima is then explored using the Monte Carlo Markov Chain algorithm {\bf emcee} \citep{Foreman2013}. To increase the performance of the modeling process, it is useful to shift the geometric origin closer to the caustics \citep{Cassan2008,Han2009a,Penny2014}. We found, like S12, that the deviations are induced by the central caustic, so we shifted the reference to the center of that caustic. $t_0$ and $u_0$ parameters are then replaced with $t_c$ and $u_c$. Note that pyLIMA follows the convention defined in \citet{Gould2004,Skowron2011}: $u_0>0$ when the lens passes the source on its right.

After finding a plausible static model, second-order effects were added to test whether they significantly improved significantly the model. It was expected that the annual parallax would be found in the event OGLE-2017-BLG-0417 due its long duration ($t_E\ge 70$ days). This can be seen in Figure~\ref{fig:static}, where the deviations in the wings of both models (after $\rm{HJD}\ge 2455840$) are typical of a parallax signature. S12 showed that including the microlensing parallax significantly improved the model. In principle, the orbital motion of the lens or the motion of the source, known as xallarap \citep{Poindexter2005,Rahvar2009,Miyake2012}, could be considered. S12 considered the former and again found a significant improvement in the likelihood of their model. In this work, we will show that only the parallax inclusion is relevant.
\subsection{Results} \label{sec:results}

\subsubsection{Static models} \label{sec:static}
As implied by the close/wide degeneracy, two competitive models exist in the wide and close binary regimes.
As can be seen in Table~\ref{tab:models}, the static wide model is significantly better than the static close model ($\Delta\chi^2\sim 400$), for the reasons we discussed in Section~\ref{sec:history}. In the absence of parallax signal, the close model trajectory passes over a triangular shape caustic. On the other hand, the wide model does not have such caustic anywhere close to the trajectory, in better agreement with the data.

\begin{figure*}    
  \centering
  \includegraphics[width=19cm]{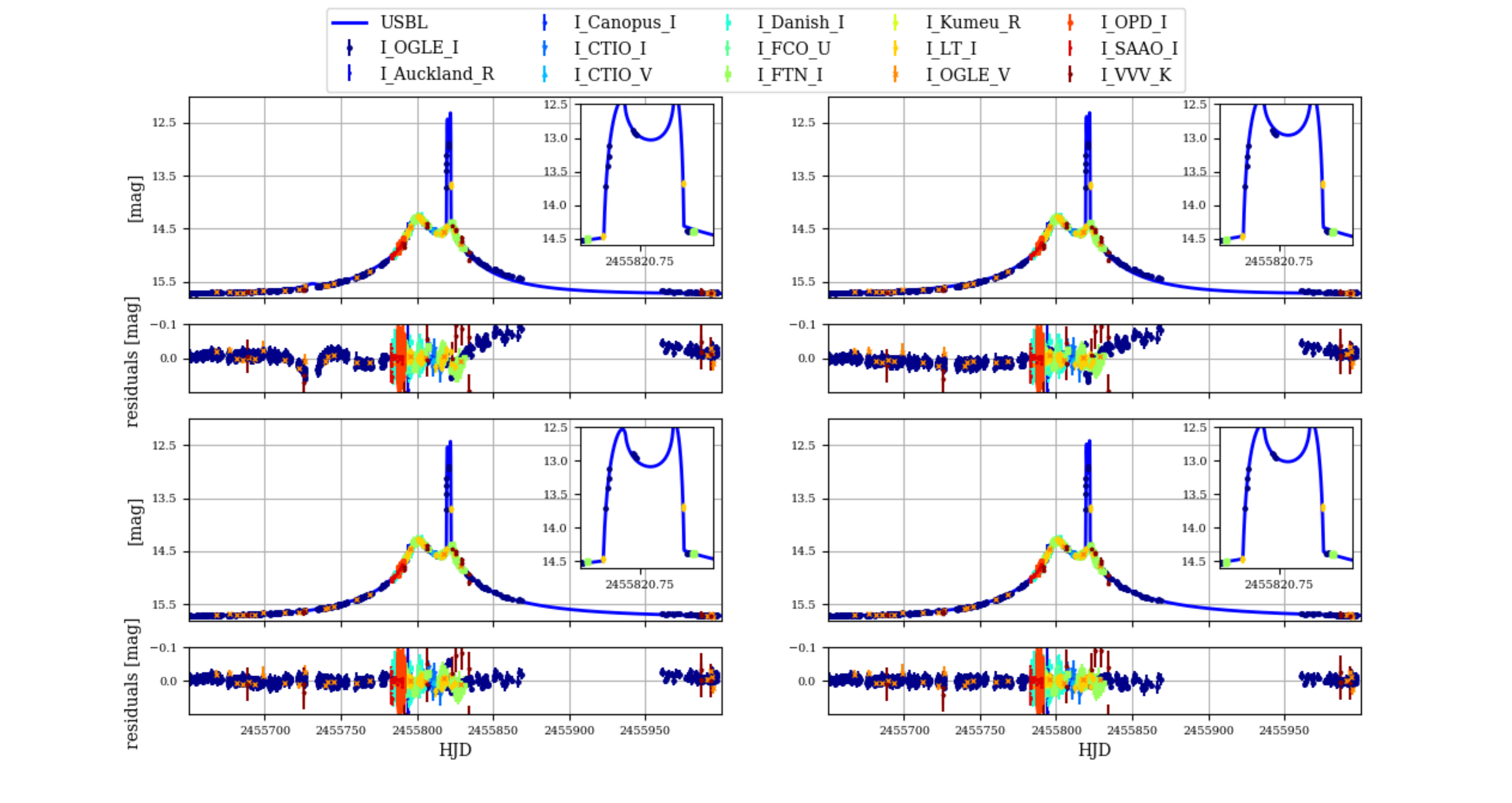}
    \caption{Best static (top) and parallax (bottom) models found in this work. The residuals for static models after 2455840 are signature of the annual parallax. VVV data are plotted but were not used for the fit. Left: Close model, note the bump due to the planetary caustic around 2455730 HJD for the static model. Right : Wide model.}
    \label{fig:static}
\end{figure*}

\begin{figure}    
  \centering
  \includegraphics[width=9cm]{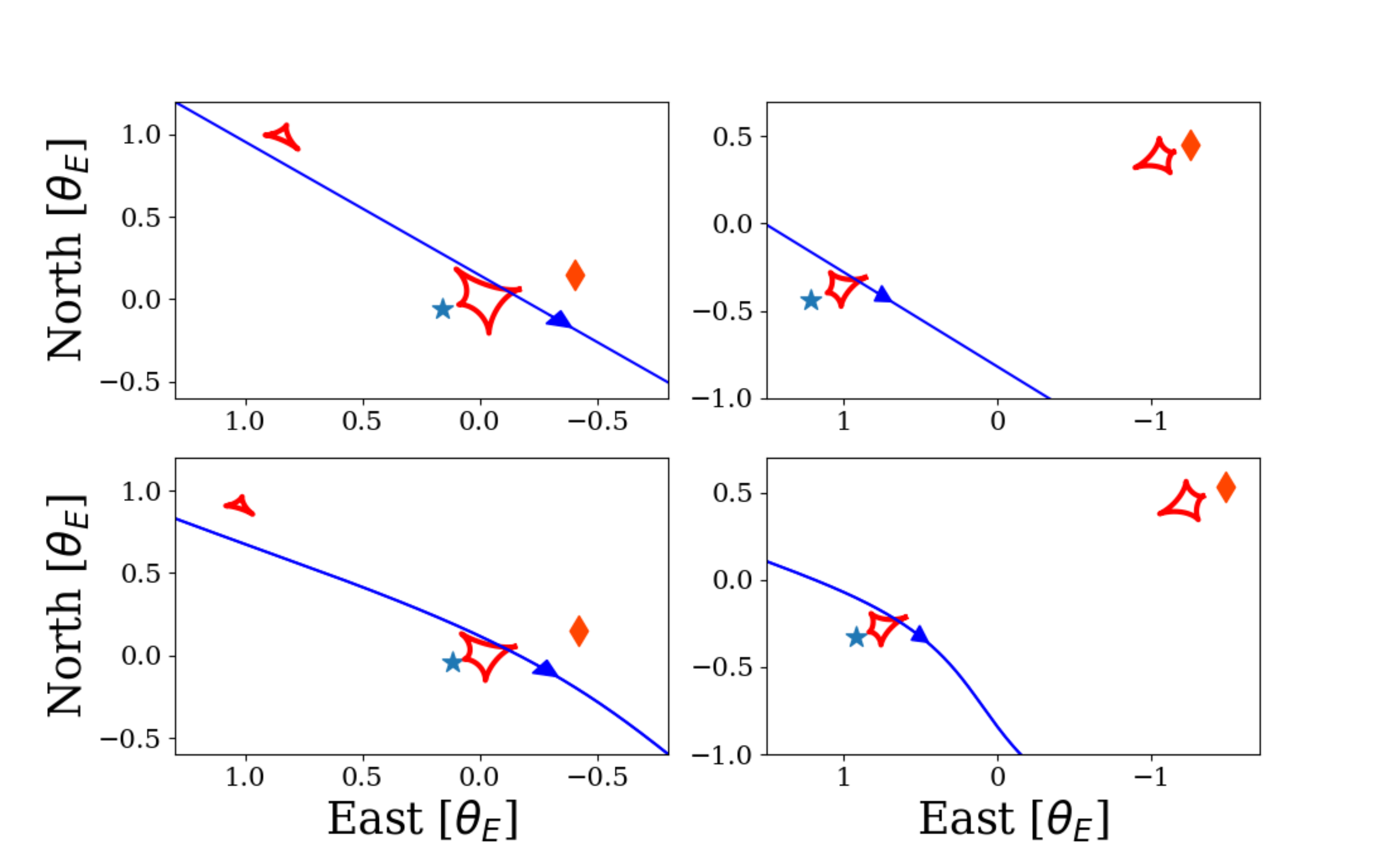}
    \caption{Caustics (red closed curves) and source trajectories (blue lines) for the static (top) and parallax ($u_c>0$) (bottom) models. Blue star and orange diamond represent the primary (heavier component) and companion of the lens, respectively. Left : Close models. Note how the parallax curvature helps to avoid the planetary caustic, reducing dramatically the $\chi^2$. Right : Wide models. In this geometry, the trajectories do not cross any planetary caustic. North is up and East is left.}
    \label{fig:caustatic}
\end{figure}

\subsubsection{Annual parallax} \label{sec:parallax}
The strong asymmetric deviations observed in the rising and falling parts of the event light curve, see Figure~\ref{fig:static}, are a classic signature of the annual parallax \citep{Gould2004}. Briefly, the orbit of the Earth around the Sun induces shifts  
in the source trajectory which significantly modifies the event magnification. Considering annual parallax effects in modeling requires two extra parameters summarized in the parallax vector, traditionally written in two different ways $\bm{\pi_E} = (\pi_{EN},\pi_{EE}) = (\pi_\parallel,\pi_\perp)$.  While the former corresponds to the projection of the parallax vector into the (North, East) plane of the sky formalism (the one used in pyLIMA), the second projects the vector in a basis parallel and perpendicular to the projected Earth acceleration at the time $t_{0,par}$.   We refer the reader to \citet{Skowron2011} for more details about the choice of $t_{0,par}$. In this work, we selected $t_{0,par}=$ 2455820 HJD, close to the caustic crossing. On this date (13 September 2011), the Earth's acceleration was nearly parallel to the vernal direction. The East component of the parallax is expected to be stronger, and therefore better constrained, because $\pi_{EE}\sim \pi_\parallel$. However, the long event duration (i.e $\ge 60$ days) should ensure a good constraint on both components. We encounter an additional degeneracy \citep{Skowron2011}, since the models are perfectly symmetric with the transformation $(u_0,\alpha,\pi_{EN}) \leftrightarrow  -(u_0,\alpha,\pi_{EN})$, known as the "ecliptic degeneracy" \citep{Skowron2011, Skowron2015}. This degeneracy is especially severe when source stars are close to the ecliptic plane, which is the case for OGLE-2011-BLG-0417 (i.e., $\lambda = 264.33^\circ$ and $\beta=-3.7969^\circ$), but this degeneracy was not investigated by S12. This degeneracy means that it is nearly impossible to distinguish the lens-source proper motion direction using only the microlensing data. Fortunately, it has only a minor impact on the mass and lens distance estimation, since they are independent of the direction (see \citet{Gould2000}, \citet{Gould2004} or Equation~\ref{eq:masses}).

\subsubsection{Orbital motion} \label{sec:om}
As underlined by S12, the orbital motion of the lens can produce significant effects on the lightcurve, especially for long-duration events \citep{Dominik1998, Skowron2011}. Moreover, it can mimic the parallax effect and therefore produce a misleading parallax measurement \citep{Batista2011, Bachelet2012a}. In this work, we implement the two-dimensional orbital motion defined by the binary lens separation expansion rate $ds/dt$ and the binary-axes rotation rate $d\alpha/dt$ \citep{Dong2009a,Batista2011}. 
To verify if the lens system is gravitationally bound, we used \citep{Dong2009a,Batista2011}:
\begin{equation}
v_\perp = \sqrt{(ds/dt)^2+s^2(d\alpha/dt)^2}D_l\theta_E \leq \sqrt{{{2GM_l}\over{s\theta_ED_l}}} ~.
\label{eq:energies}
\end{equation}
We found that including the orbital motion significantly improved the close model interpretation (with $\Delta\chi^2\sim 600$, see Table~\ref{tab:models}). The close model with orbital motion reported in this work is bound since $v_\perp = 4.4 ~ \leq 7.7 ~\rm{au}~ \rm{yr^{-1}}$. This model is similar to that found by S12 and as suspected in Section~\ref{sec:history}, the triangular caustic is located far from the source trajectory at $\rm{HJD}=2455730$ (at this date, the triangular caustic is located at $(x,y) = (-2,3)$). This allows the model to tune the parameters to fit the residuals remaining in the close parallax fit visible in the Figure~\ref{fig:static}. This effect also impacts the fit parameters $t_E$, $\pi_{EN}$ and $\pi_{EE}$: adding the orbital motion effect changes their values significantly.

Conversely, the orbital motion effect is not measured for the wide models. Both the likelihood improvement and orbital motion parameters values are negligible. This is not surprising, since no sign of systematics remains in the residuals of the wide parallax model in the Figure~\ref{fig:static}. Moreover, $t_E$, $\pi_{EN}$ and $\pi_{EE}$ stayed consistent with or without the inclusion of the orbital motion effect.

\subsubsection{Model selection} \label{sec:selection}
Statistical tools exist for optimized model selection, provided that their parameters are linear (see for example \citet{Bramich2016} for a review). However, these methods cannot be used in the present case because the model parameters are non linear. Instead, we adopt the following argument, by applying the Occam's Razor. First, the wide models reproduce the data better than the close models, with or without parallax taken into account. Moreover, the parallax wide models found in this work requires fewer paramaters than the 3D orbital motion model with a similar likelihood. Table~\ref{tab:models} shows that orbital motion is not detected for wide models. The most plausible models are therefore the wide parallax models.

For the rest of this work, we will then refer to the positive ($u_c>0$) wide parallax model as the best model, but similar conclusions can be made using the negative ($u_c<0$) wide parallax model. In Section~\ref{sec:propmotion}, we discuss how future observations can distinguish these two models.

\section{Reassemble the puzzle} \label{sec:puzzle}
\subsection{Source and blend brightness} \label{sec:mags}
\begin{turnpage}

\begin{table*}
  \centering
   \scriptsize
  \begin{tabular}{lccccccccccc}
     
     \hline\hline
     \multirow{2}{*}{Parameters} &
      \multicolumn{3}{c}{S12} &
      \multicolumn{3}{c}{pyLIMA close} &
      \multicolumn{5}{c}{pyLIMA wide} \\
& Standard & Parallax & Orbital Motion & Standard & Parallax ($u_c>0$) & Orbital Motion ($u_c>0$) & Standard & Parallax ($u_c>0$) & Parallax ($u_c<0$) & Orbital Motion ($u_c>0$) & Orbital Motion ($u_c<0$) \\
\hline
&&\\
$t_{ref}-2450000~[HJD]$ & 5817.302(18)&5815.867(30)&5813.306(59)&5816.387(27)&5814.785(60)&5812.10(17)&5811.517(29)&5810.92(12)&5811.05(14)&5811.37(52)&5811.12(53) \\
$u_{ref}$ & 0.1125(1) & -0.0971(3) & -0.0992(5) & 0.13608(16)&0.11659(55)&0.1148(17)&0.11124(63)&0.11071(88)&-0.1128(11)&0.1136(39)&-0.1147(37) \\
$t_E~[days]$ & 60.74(8) & 79.59(36) & 92.26(37) & 60.79(11)&80.97(56)&99.5(1.6)&98.23(52)&105.79(62)&104.27(70)&103.5(2.5)&105.7(2.7) \\ 
$\rho~(10^{-3})$ & 3.17(1) & 2.38(2) & 2.29(2) &3.094(34) & 2.494(22)&2.127(59)&2.138(21)&2.156(37)&2.202(43)&2.179(50)&2.219(48) \\
$s_\perp$ & 0.601(1) & 0.574(1) & 0.577(1) & 0.60184(39) &0.5731(12)&0.5681(32)& 2.6214(66)&2.5440(69)&2.5374(62)&2.5378(98)&2.5318(94) \\
$q$ & 0.402(2) & 0.287(2) & 0.292(2) & 0.3967(26)&0.2829(27)&0.2800(40)&0.968(12)&0.617(10)&0.629(12)&0.619(12)&0.627(11) \\
$\alpha~[\rm{rad}]$ & 1.030(2) & -0.951(2) & -0.850(4) &-1.0257(20)&-0.9609(30)&-0.8994(71)& -0.8422(17)&-0.8486(19)&0.8575(88)&-0.875(35)&0.875(33) \\
$\pi_{EN}$ &* & 0.125(4) &0.375(15) &*&-0.1263(56)&-0.314(30)&*&-0.137(19)&0.175(25)&-0.128(21)&0.189(25)\\
$\pi_{EE}$ & * & -0.111(5) &-0.133(3)&*&-0.1151(65)&-0.1500(56)&*&-0.1603(32)&-0.1495(44)&-0.1576(60)&-0.1432(66)\\
$ds_\perp/dt~[yr^{-1}]$ & * & * &1.314(23) &*&*&1.353(48)&*&*&*&0.06(17)&0.20(15)\\
$d\alpha/dt~[\rm{rad} ~\rm{ yr^{-1}}]$ & * & * &1.168(76) &*&*&1.26(25)&*&*&*&-0.12(13)&-0.03(0.14)\\
$s_\parallel$ & * & * &0.467(20) &*&*&*&*&*&*&*&*\\
$ds_\parallel/dt~[\rm{yr^{-1}}]$ & * & * & -0.192(36) &*&*&*&*&*&*&*&*\\
$\chi^2/\rm{dof}$ & 4415/2627 & 2391/2625 &1735/2621&4456/2627&2348/2625&1741/2623&4026/2627&1723/2625&1724/2625&1720/2621&1720/2621\\
&&\\
  \hline 
  \end{tabular}

  \caption{Summary of the various models presented in this work. Numbers in brackets are the $1\sigma$ uncertainties. The columns corresponding to pyLIMA report the best model while uncertainties are derived from the MCMC exploration (i.e 68\% intervals). Note that $t_{ref}$ and $u_{ref}$ refers to $t_0$ and $u_0$ for S12 models and $t_c$ and $u_c$ for pyLIMA models.}
  \label{tab:models}
\end{table*}
\end{turnpage}
Using our MCMC exploration around the best model (i.e the $u_c \ge 0$ wide parallax model), we obtain 
the relative brightness of the source and the blend in several bands. For the source, we found 
$I_s = 16.47 \pm 0.01$ mag, $V_s = 19.04 \pm 0.01$ mag. The brightness of the blend are $I_b = 16.52 \pm 0.01$ mag, $V_b = 18.104 \pm 0.005$ mag.
The $I$ and $V$ bands are in the OGLE-IV system, very close to the Johnson-Cousin system \citep{Udalski2015a}, whereas S16 assumed the AB system.
G13 derived a fainter source $(V,I)_s=(19.42, 16.73)$ mag and a brighter blend $(V,I)_l=(18.23,16.29)$ mag, in good ($I$-band) and relative ($V$-band) agreement with our close parallax model estimations: $(V,I)_s=(19.28,16.71)$ mag and $(V,I)_b=(18.05,16.31)$ mag.

The Interstellar Extinction Calculator on the OGLE website\footnote{http:http://ogle.astrouw.edu.pl/}, based 
on \citet{Nataf2012} and \citet{Gonzalez2012}, returns $A_I =2.0 \pm 0.1$ mag (we assume the error on $A_I$) and $E(V-I)=1.58 \pm 0.08$ mag. 
The absorption corrected brightness of the source is then $I_{o,s} = 14.5 \pm 0.1$ mag and the intrinsic color is $(V-I)_{o,s} = 1.0 \pm 0.1$ mag. Finally, the color 
needs to be corrected by a factor 0.94 \citep{Udalski2015a} (the microlensing event was on the CCD 22 of the OGLE-IV mosaic camera) to be in the standard system, leading to a final color $(V-I)_{o,s} = 0.94 \pm 0.1$ mag. Using the color-magnitude-radius relation of \citet{Kervella2008}, we found $\theta_*= 5.0 \pm 0.3 ~\rm{\mu as}$ and finally $\theta_E = 2.3 \pm 0.2 $ mas and $\mu_{geo}=8.0 \pm0.8 ~\rm{mas~yr^{-1}}$ (in the geocentric frame).

Thanks to VVV observations, it is possible to estimate the source flux in the near infrared. However, because the target was so bright during the peak magnification (i.e., $\le 12$ mag) that the target approached saturation. We take this into account by weighting the brightness estimations with a 0.2 mag error. Using the MCMC exploration, we compute the magnification relative to the VVV observations and estimate a source brightness of $K_s=13.47 \pm 0.2$ mag. We can estimate the near infrared absorption based on the extinction maps of \citet{Gonzalez2012}. Using their online tool\footnote{http://mill.astro.puc.cl/BEAM/calculator.php},
we found (with a box size of 2' and using the \citep{Nishiyama2009} extinction law) $A_K=0.3\pm0.1$ mag and $E(J-K)=0.6\pm0.1$ mag. The source brightness is therefore $K_{o,s} = 13.1 \pm 0.1$ mag and $(I-K)_{o,s} = 1.3 \pm 0.1$ mag. 

Both colors are in good agreement with a G8 giant source \citep{Bessell1988}. S16 have obtained a high-resolution image in the Ks-band and found the target to be $K=12.94 \pm 0.03$ mag. They did not find any significant companion to the target star, indicating that the blend and the source are still aligned. This is consistent with our proper motion measurement. This image was taken 4 years after the event peak, so we would expect a separation of $\sim30$ mas, equivalent to 3 pixels of the narrow camera of NIRC2. So assuming the object they observed is the combined image of the source and the lens, we estimate that the lens brightness is $K_b=13.97 \pm 0.2$ mag.

\subsection{Is the blend the lens?} \label{sec:blendlens}
In common with S12 and G13, we test the hypothesis that the blend could be the lens using several evidences. First, using the parallax norm $\pi_E = \sqrt{\pi_{EN}^2+\pi_{EE}^2}$ and the angular Einstein ring radius, we can derive the total mass of the lens \citep{Gould2000}:
\begin{equation}
M_{tot} = {{\theta_E^2}\over{\kappa\pi_{rel}}} = {{\pi_{rel}}\over{\kappa\pi_E^2}}={{\theta_E}\over{\kappa\pi_{E}}}
\label{eq:masses}
\end{equation}
with $\pi_{rel}=\rm{au}(D_l^{-1}-D_s^{-1})$. 

We can combine the brightness measurements of the blend derived previously with isochrones from \citet{Bressan2012} and \citet{Marigo2017}  \footnote{http://stev.oapd.inaf.it/cgi-bin/cmd} (with solar metallicity and 5 Gyr isochrones) to estimate the total mass and distance of the blend under two assumptions. We assume the blend to be a binary star with the same mass ratio as the lens (i.e., $q=0.61$) and use the partial extinction law of \citet{Bennett2015} and \citet{Beaulieu2016}. 

Finally, we take advantage of the recent GAIA data release \citep{GAIA2016,GAIA2018} and found the target parallax : $p=0.8 \pm 0.2$ mas (the GAIA DR2 ID is 4061372412899533056). This leads directly to a target distance of $D_l = 1.3 \pm 0.3$ kpc.  The significant blending $g=f_b/f_s\sim1$ of this event affects the parallax measurement because GAIA has measured the displacement of the photocenter in this case. We therefore used a refined estimate of the distance, $D_l\in[982,1734]$ pc, available in the catalog of \citet{BailerJones2018}.

\begin{figure}    
  \centering
  \includegraphics[width=9cm]{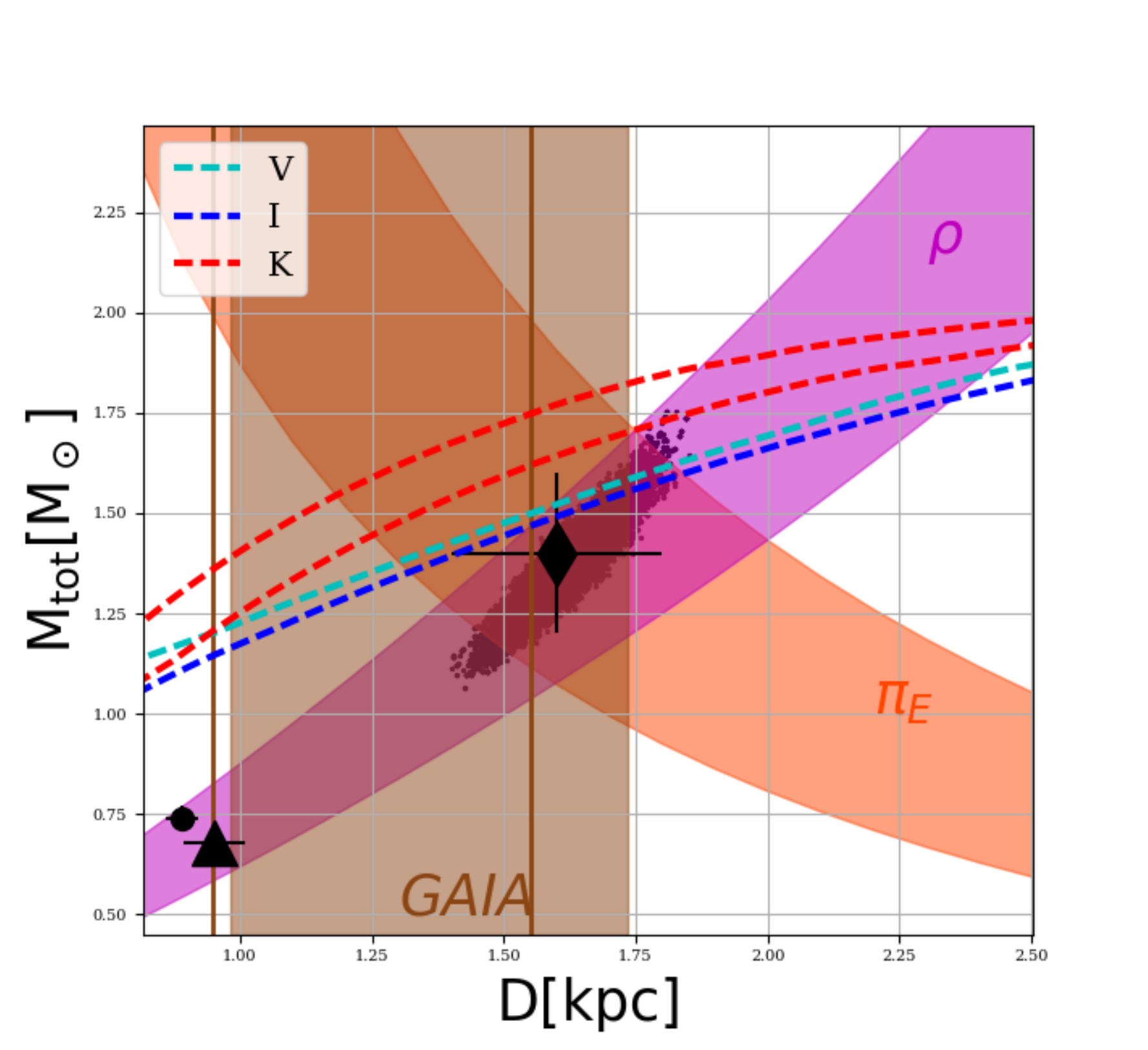}
    \caption{Various constraints for the lens mass and distance (assuming the blend is the lens). Black dots are results from the MCMC exploration around the best model (Model wide parallax $u_c>0$). The black points with errorbars are solutions of S12 (circle), G13 (triangle) and this work (diamond). The magenta area represents the constraint from the source size measurement (i.e., $\rho$). The orange shaded area represents the microlensing annual parallax constraint. The brown shaded region indicates the GAIA parallax constraints from \citet{BailerJones2018} (the two vertical brown lines indicates the "brute" GAIA distance) and the three colored dashed curves represent the isochrones of \citet{Bressan2012}, assuming a 5 Gyr binary system of mass ratio $q=0.61$ for the blend. We plot the $1 \sigma$ region for the $K$-band lens magnitude due to the large uncertainties.}
    \label{fig:massdistance}
\end{figure}

We combine this information in Figure~\ref{fig:massdistance}. It is clear that the blend is likely to be the lens. All constraints converge to a total lens mass $M_{tot} = 1.4 \pm 0.2 ~\rm{M_\odot}$ and $D_l=1.6 \pm0.2$ kpc. According to this scenario, the lens is a binary composed of a K-dwarf with $M_{l,1} = 0.9 \pm 0.1~\rm{M_\odot}$ and an M-dwarf with $M_{l,2} = 0.5 \pm 0.1~\rm{M_\odot}$.

\subsection{Wide scenario predictions versus B15 and S16 measurements} \label{sec:previousconstrains}

Following G13, we make a prediction of the radial velocity that would be expected from follow-up observations, based on this new model. Using our best model, we find that the orbital radius of the binary is at least $a\sim 16$ au, leading to a minimum orbital period $P\sim 52$ yr. According to this new scenario, it is not surprising that B15 did not find any signal in their observations. 

The wide model confirms the predictions by S16 that the source is redder ($(V-I)_s = 2.57$ mag) than the lens ($(V-I)_b=1.58$ mag). The source brightnesses derived in this work are in agreement with the spectral energy distribution (SED) study of S16 (this work, S16): $V=(19.04\pm 0.01,19.4\pm 0.58)$ mag, $I=(16.47\pm 0.01,16.30\pm 0.34)$ mag, $K=(13.47\pm 0.2,13.55 \pm 0.17)$ mag. Note that we remodel the SED presented in S16, but using a scenario with a giant source and two bright lenses. While the source and the lens primary properties stay unchanged, the mass of the companion is weakly constrained ($M_{l,2} = 0.24^{+0.27}_{-0.11}~\rm{M_\odot}$) but in agreement with the estimation made in the previous section.
Given the masses of the individual components ($M_{l,1} = 0.9 \pm 0.1~\rm{M_\odot}$ and $M_{l,2} = 0.5 \pm 0.1~\rm{M_\odot}$), one can expect that the flux ratios between the binary components to be $\ge 4$ in the wavelength windows observed by S16. The SED constraints on the companion are therefore expected to be weak.

\subsection{Predictions with GAIA DR2 proper motions} \label{sec:propmotion}
In addition to the parallax information, the GAIA DR2 also provides proper motion measurements in the ICRS reference frame $\bm{{\mu_{G}}}(E,N) = (-4.06 \pm 0.44, -6.39 \pm 0.35)~\rm{mas~yr^{-1}}$. Note that the components of the GAIA DR2 proper motions are $\mu_{\alpha ^*} = \mu_{\alpha}\cos(\delta)$ and $\mu_\delta$ \citep{Luri2018}. This measurement is also affected by the blending and, assuming the blend light is coming only from the lens, the reported proper motion is the proper motion of the photocenter, similarly to the astrometric microlensing \citep{Dominik2000,Nucita2017}:
\begin{equation}
\bm{\mu_{G}} = {{f_s\bm{\mu_{s}}+f_b\bm{\mu_{l}}}\over{f_s+f_b}} = {{g\bm{\mu_{l}}+\bm{\mu_{s}}}\over{1+g}}~,
\label{eq:muphoto}
\end{equation}
with $\bm{\mu_{l}}$ and $\bm{\mu_{s}}$ representing the proper motion of the lens and the source respectively. The microlensing models predict the heliocentric lens-source proper motion \citep{Skowron2011}:
\begin{equation}
\bm{\mu_{helio}} = \bm{\mu_{geo}}+\pi_{rel}\bm{V_{\oplus,\perp}}~,
\label{eq:muhelio}
\end{equation}
where $\bm{V_{\oplus,\perp}} = (-1.99,0.58) \rm ~ km~s^{-1}$ is the projected velocity of the Earth at the time $t_{0,par}$ \citep{Gould2004}. The proper motion vector is given by :
\begin{equation}
\bm{\mu_{geo}} = {{\bm{\pi_E}}\over{\pi_E}}\mu_{geo}~.
\label{eq:mugeo}
\end{equation}
Knowing that :
\begin{equation}
\bm{\mu_{helio}} = \bm{\mu_{l}}-\bm{\mu_{s}} ~,
\end{equation}
one can rewrite Equation~\ref{eq:muphoto} and derives:
\begin{equation}
\bm{\mu_{l}} =\bm{\mu_{G}} + {{\bm{\mu_{helio}}}\over{1+g}}
\label{eq:propsolve}
\end{equation}
and
\begin{equation}
\bm{\mu_{s}} =\bm{\mu_{G}} - g{{\bm{\mu_{helio}}}\over{1+g}} ~.
\label{eq:propsolve}
\end{equation}

Assuming that the blend light is measured and emitted by the lens, the proper motions of the source and the lens can be estimated.
For the event OGLE-2011-BLG-0417, the blending of the source is $g\sim$ 1. Note that it is difficult to precisely estimate the blending for the GAIA G band \citep{Jordi2010} because the filter is broad and, more importantly, there is no available lightcurve. As can be seen in Table~\ref{tab:mu}, the projected speeds of the lens and the source can be estimated assuming (for both scenarios) a source distance of $D_s=8.2 \pm 0.5$ kpc, a lens distance $D_l=1.6 \pm 0.2$ kpc, $\theta_E=2.3 \pm 0.2$ mas, $\mu_{geo} = 8 \pm 0.8 ~\rm{mas~yr^{-1}}$ and $g=1 \pm 0.1$. Details about the estimation of errors can be found in Appendix~\ref{sec:coordtransformation}.

\begin{figure}    
  \centering
  \includegraphics[width=9cm]{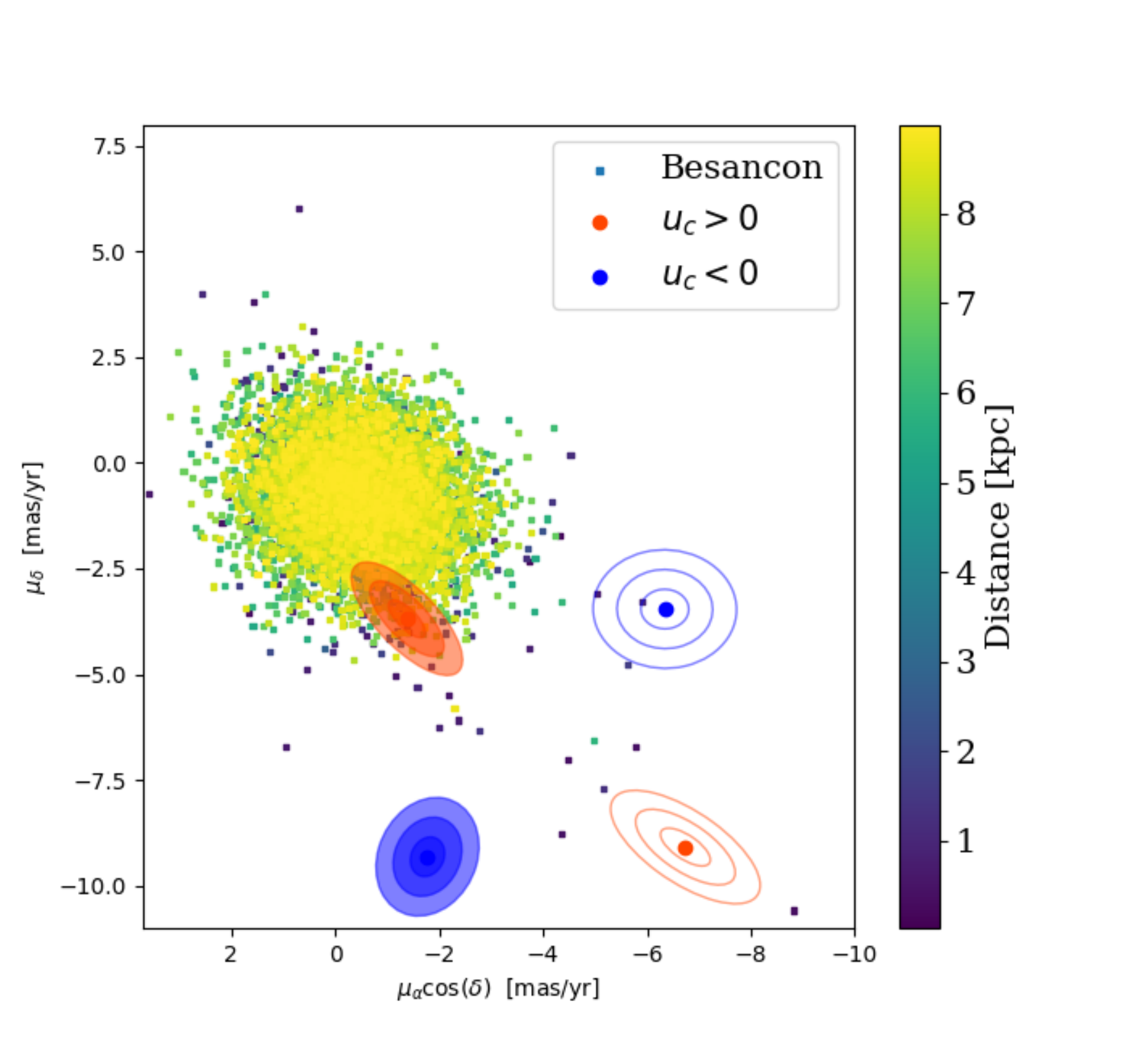}
    \caption{Proper motions in equatorial coordinates. The Besancon model predictions are represented with small squares, and the colors indicate the distance from the Earth. Solid (open) ellipses represent the sources (lenses) proper motions, orange (blue) for the $u_c>0$ ($u_c<0$) models.}
    \label{fig:mu}
\end{figure}

Using the coordinate system transformation detailed in Appendix~\ref{sec:coordtransformation}, the proper motions vectors of the source in galactic coordinates are $(\mu_{l},\mu_{b}) = (-3.85\pm 0.53,-0.83 \pm 0.21)~\rm{mas~yr^{-1}}$ ($u_c>0$) and $(-8.82\pm 0.40,-3.54 \pm 0.41~\rm{mas~yr^{-1}}$)  ($u_c<0)$ for the wide parallax models. Both models therefore predict that the source moves against the rotation of the Galactic Disk. It seems this is also the case for the lens, since the proper motions vectors of the lens are $(-11.30\pm 0.53,0.83 \pm 0.21)~\rm{mas~yr^{-1}}$ ($u_c>0$) and $(-6.34\pm 0.40, 3.46 \pm 0.41)~\rm{mas~yr^{-1}}$ ($u_c<0$). Figure~\ref{fig:mu} compares the results from the Table~\ref{tab:mu} with the Besancon galactic model predictions\footnote{http://model.obs-besancon.fr/} \citep{Robin2004}. While it is impossible to distinguish the degenerate models from the microlensing predictions, or the radial velocity observations, the proper motion analysis from the GAIA DR2 catalog seems to significantly favor the $u_c>0$ model. This conclusion can be tested in the near future (i.e $\sim$ 5 yr) by comparing S16 observations and new high resolution imaging. There is no ambiguity between the two model predictions, since the opposite direction of the North components of the heliocentric proper motions seen in Table~\ref{tab:mu} is significantly detected.

\begin{table*}
  \centering
       \scriptsize
  \begin{tabular}{lccccc}

     \hline\hline
   
Model & $\bm{\mu_{helio}}~ \rm mas~yr^{-1}$ &  $\bm{\mu_{s}}~ \rm mas~yr^{-1}$ & $\bm{V_{s,\perp}}~\rm km~s^{-1}$ & $\bm{\mu_{l}}~\rm mas~yr^{-1}$ &$\bm{V_{l,\perp}} ~\rm km~s^{-1}$ \\
\hline
&&\\
Wide parallax $u_c>0$ & $(-6.02 \pm 0.62, -5.40 \pm 0.55)$ & $(-1.55 \pm 0.40,-3.69 \pm 0.45)$&$(-60\pm 16,-143\pm 20)$& $(-7.57\pm 0.40,-9.09\pm 0.45)$& $(-57.4\pm 7.8,-69.0\pm 9.3)$ \\
Wide parallax $u_c<0$ & $(-5.13 \pm 0.54, 5.86 \pm 0.62)$ &  $(-2.00 \pm 0.37,-9.32 \pm 0.47 )$&$(-77\pm 15,-362 \pm 29)$& $(-7.13\pm 0.37,-3.46 \pm 0.47)$& $(-54.1\pm 7.3,- 26.2\pm 4.9)$ \\
\\
  \hline 
  \end{tabular}
  \caption{Heliocentric proper motion and speed estimations for OGLE-2011-BLG-0417 (East, North). Proper motions are in the system $\mu=(\mu_\alpha,\mu_\delta)$. Note that the errors of $\bm{\mu_{l}}$ and $\bm{\mu_{s}}$ are strictly equal since $g=1$.}
  \label{tab:mu}
\end{table*}
 
\section{Discussion and conclusion} \label{sec:conclusion}
From Figure~\ref{fig:massdistance}, we notice that the $K$-band constraint is not in good agreement with the others. This may be explained by the extreme brightness of the event in this band, potentially leading to an underestimation of the source flux in $K$ (and then an overestimation of the lens brightness). The isochrones, especially for M-stars, are also sensitive to the age, metallicity and potentially dust models (whereas we have not considered dust in this work) \citep{Marigo2017}.   

Besides the $K$-band brightness constraint, all constraints converge to a total lens mass $M_{tot} = 1.4 \pm 0.2~ \rm{M_\odot}$ and $D_l=1.6 \pm0.2$ kpc. According to this new scenario, the lens is a binary composed of a K-dwarf with $M_{l,1} = 0.9 \pm 0.1~\rm{M_\odot}$ and an M-dwarf with $M_{l,2} = 0.5 \pm 0.1~\rm{M_\odot}$. The lens orbital radius of the binary is at least $a\ge 16$ au with a minimum orbital period $P\ge 52$ yr, contrary to the conclusion of G13. The absence of radial velocity modulations measured in B15 is in agreement with this new scenario. More generally, our magnitude predictions for the source (but not the blend) are in agreement with S16.

We found two degenerate models in the wide-binary regime caused by the symmetry of the source trajectory with respect to the binary lens axis. We present a method using the GAIA proper motion measurements, and predict that the model $u_c>0$ is preferred. This is verifiable with future high-resolution imaging.

Since there has not yet been developed a method to decide the optimal range and spacing in grid search for lensing parameters, it is highly recommended to cross check results from this technique using an alternative method to explore parameters space. This especially true for models that require extreme tuning of the source trajectory (i.e high orbital motion and/or parallax and/or xallarap). As the grid method is computationally intensive, applying it to large datasets results in excessive analysis timescales and requires costly many-CPU computing clusters. It is therefore beneficial and timely to explore alternative that can handle microlensing datasets expected from the planned missions WFIRST and Euclid. 
\section*{Acknowledgements}
The authors thank the anonymous referee for the constructive comments. The authors thank In-Gu Shin for the access to the data and details about the original publication.
Work by EB and RAS is support by the NASA grant NNX15AC97G.
JPB was supported by the University of Tasmania through the UTAS Foundation and the endowed Warren Chair in Astronomy,
the CNES and INSU PNP.
This research has made use of NASA's Astrophysics Data System. $\mu=(\mu_\alpha,\mu_\delta)$
This work has made use of data from the European Space Agency (ESA)
mission {\it Gaia} (\url{https://www.cosmos.esa.int/gaia}), processed by
the {\it Gaia} Data Processing and Analysis Consortium (DPAC,
\url{https://www.cosmos.esa.int/web/gaia/dpac/consortium}). Funding
for the DPAC has been provided by national institutions, in particular
the institutions participating in the {\it Gaia} Multilateral Agreement.
This publication makes use of data products from the Two Micron All Sky Survey, which is a joint project of the University of Massachusetts and the Infrared Processing and Analysis Center/California Institute of Technology, funded by the National Aeronautics and Space Administration and the National Science Foundation. This research made use of Astropy, a community-developed core Python package for Astronomy (Astropy Collaboration, 2013). This research has made use of the SIMBAD database,
operated at CDS, Strasbourg, France. This work has made use of data from the European Space Agency (ESA)
mission {\it Gaia} (\url{https://www.cosmos.esa.int/gaia}), processed by
the {\it Gaia} Data Processing and Analysis Consortium (DPAC,
\url{https://www.cosmos.esa.int/web/gaia/dpac/consortium}). Funding
for the DPAC has been provided by national institutions, in particular
the institutions participating in the {\it Gaia} Multilateral Agreement.
Based on observations made with ESO Telescope at the Paranal
Observatory under program ID 092.C-0763(A) and 093.C-0532(A).
Based on observations at Cerro Tololo Inter-American Observatory,
National Optical Astronomy Observatory, which is operated by the As-
sociation of Universities for Research in Astronomy (AURA) under a
cooperative agreement with the National Science Foundation.

\appendix

\section{Propagation of uncertainties in proper motion transformations} \label{sec:coordtransformation}

Our goal is to derive the proper motion of the lens and the source in the galactic and equatorial systems. The transformation of proper motions from equatorial coordinates $\mu_{\alpha *}$ and $\mu_{\delta}$ to galactic coordinates $\mu_{l *}=\mu_{l}\cos(b)$ and $\mu_{b}$ is given by a rotation matrix \citep{Johnson1987,Luri2018, Poleski2013}:
\begin{equation}
\begin{pmatrix}
\mu_{l*}\\
\mu_{b}
\end{pmatrix} = R
\begin{pmatrix}
\mu_{\alpha *}\\
\mu_{\delta}
\end{pmatrix}
\end{equation}

Differentiating the galactic coordinates defined in \citep{Binney1998}, one can derive \citep{Poleski2013} :
\begin{equation}
R = {{1}\over{\cos(b)}} 
\begin{pmatrix}
R_1 & R_2 \\
-R_2 & R_1
\end{pmatrix}
\label{eq:rotation}
\end{equation}
with
\begin{equation}
R_1 = \cos(\delta)\sin(\delta_{GP})-\sin(\delta)\cos(\alpha-\alpha_{GP})\cos(\delta_{GP})~;~R_2 = \sin(\delta_{GP})\sin(\alpha-\alpha_{GP})
\end{equation}
Since the rotation matrix determinant has to be unity, it implies $\cos(b)=\sqrt{R_1^2+R_2^2}$. 

 \citet{Luri2018} stress that the correlation terms can seriously impact the estimation of errors in the proper motion. Considering the general problem $Y=f(X)$, the variance-covariance matrices relation is \citep{Luri2018}:
\begin{equation}
C_Y = JC_XJ^T
\end{equation}
where $C_Y$ and $C_X$ are the variance-covariance matrices and $J$ the Jacobian matrix of the transformation. In the following, we describe the various components of the error estimations of quantities define in Section~\ref{sec:propmotion} in the $\mu=(\mu_\alpha,\mu_\delta)$ system. First, the variance-covariance $C_{\mu_{geo}}$ matrix of the vector $\bm{\mu_{geo}}$ define in Eq~\ref{eq:mugeo} is :
\begin{equation}
C_{\mu_{geo}} = J
\begin{pmatrix}
A&B \\
B^T&C_{\pi_E}
\end{pmatrix}
J^T~;~ J=
{{1}\over{\pi_E}}\begin{pmatrix}
\pi_{EE} & \mu_{geo}\pi_{EN}^2 & \mu_{geo}\pi_{EE}^2  \\
\pi_{EN} & \mu_{geo}\pi_{EE}^2 & \mu_{geo}\pi_{EN}^2 
\end{pmatrix}
\end{equation}
where $C_{\pi_E}$ is the variance-covariance matrix of the normalised microlensing parallax, $A=\sigma_{\mu_{geo}}^2$ (i.e the scalar error) and $B=(0,0)$. The former is estimated numerically from the MCMC exploration. Similarly, we can derive the $C_{\mu_{helio}}$ matrix:
\begin{equation}
C_{\mu_{helio}} = J
\begin{pmatrix}
C_{\mu_{geo}}&B^T \\
B&C_{\pi_E}
\end{pmatrix}
J^T~;~ J=
\begin{pmatrix}
1 & 0 & V_{E,E}  \\
0 & 1 & V_{E,N}
\end{pmatrix}
\end{equation}
The errors of the source proper motion can be estimated from:
\begin{equation}
C_{\mu_{s}} = J
\begin{pmatrix}
C_{\mu_{G}}&D&B^T \\
D&C_{\mu_{helio}}&B^T \\
B&B&\sigma_g^2 \\
\end{pmatrix}
J^T~;~ J=
\begin{pmatrix}
1 & 0 & -{{1}\over{1+g}}&0&{{1}\over{(1+g)^2}} \\
0 & 1 & 0 &-{{1}\over{1+g}}&{{1}\over{(1+g)^2}} \\
\end{pmatrix}
\end{equation}
with $\sigma_g$ is the error on the blending ratio $g=f_b/f_s$ between the lens and the source and $C_{\mu_{G}}$ is the variance-covariance matrix of GAIA proper motions (estimated from the GAIA DR2 archive and transformed to the $\mu=(\mu_\alpha,\mu_\delta)$ system). The matrix $C_{\mu_{l}}$ is:
\begin{equation}
C_{\mu_{l}} = J
\begin{pmatrix}
C_{\mu_{G}}&D&B^T \\
D&C_{\mu_{helio}}&B^T \\
B&B&\sigma_g^2 \\
\end{pmatrix}
J^T~;~ J=
\begin{pmatrix}
1 & 0 & {-{g}\over{1+g}}&0&-{{1}\over{(1+g)^2}} \\
0 & 1 & 0 &{-{g}\over{1+g}}&-{{1}\over{(1+g)^2}} \\
\end{pmatrix}
\end{equation}
Note that the errors for $\bm{\mu_{l}}$ and $\bm{\mu_{s}}$ are therefore nearly similar. In Section~\ref{sec:propmotion}, they are strictly identical since $g=1$. The errors for the source and lens speeds can be estimated using:
\begin{equation}
C_{V_{i}} = J
\begin{pmatrix}
C_{\mu_{i}}&B^T \\
B&\sigma_{D_i}^2 
\end{pmatrix}
J^T~;~ J=
\begin{pmatrix}
D_i & 0 & \mu_{i,E} \\
0 & D_i & \mu_{i,N}
\end{pmatrix}
\end{equation}
and the index $i$ indicates the source or the lens respectively.
The transformation to the system $\mu_*=(\mu_\alpha\cos(\delta),\mu_\delta)$ of any matrix $C$ is simply:
\begin{equation}
C_* = JCJ^T~;~ J=
\begin{pmatrix}
-\sin(\delta)&0 \\
0&1
\end{pmatrix}
\end{equation}
The errors from equatorial to galactic coordinates is given by \citep{Luri2018}:
\begin{equation}
C_{g*} = RC_*R^T
\end{equation}
since the Jacobian of a rotation is the rotation itself. Finally the transformation to the system $(\mu_l,\mu_b)$ from the system $(\mu_{l*}=\mu_l\cos(b),\mu_b)$ is:
\begin{equation}
C_g = JC_{g*}J^T~;~ J=
\begin{pmatrix}
-\sin(b)/\cos(b)^2&0 \\
0&1
\end{pmatrix}
\end{equation}
We use in this study the updated coordinates of the North Galactic Pole $\alpha_{GP}=192.85948$ and $\delta_{GP}=27.1285$ \citep{Perryman1997, Padmanabhan2002, Poleski2013}.
\bibliographystyle{aa}
\bibliography{biblio_ob150417.bib}

\end{document}